# High-Performance NC for HSM by means of Polynomial Trajectories

C. Lartigue[1,2], C. Tournier[1], M. Ritou[1], D. Dumur[3](1)
[1] LURPA-ENS de Cachan-Université Paris XI, 61 av Pt Wilson, 94235 Cachan - France
[2] IUT de Cachan-Université Paris XI, 9 av Division Leclerc, 94234 Cachan – France
[3] SUPÉLEC, Plateau de Moulon, F 91 192 Gif sur Yvette cedex – France

**Abstract**
This paper summarises works carried out for defining tool trajectory formats well adapted to High Speed Machining (HSM). Advantages in using native polynomial formats, calculated directly from the CAD model, are highlighted. In particular, polynomial surface formats are presented as a generic format for tool trajectory. Illustrations show that surface formats represent a good compromise between smoothness machining time, and surface quality.

## 1 INTRODUCTION

High-Speed Machining (HSM) is now widely used for the machining of sculptured surfaces as a high level of geometrical surface quality can be reached within a competitive machining time. Machining of sculptured surfaces relies on the calculation of tool trajectories defining the motion of the tool. The CAM system calculates the tool trajectory and the related feedrates, information which is afterward transmitted to the Numerical Controller (NC) unit. The role of the NC is thus to control the machine tool axes to move the tool at the specified feedrate, while respecting a maximum contour error lag. As a result, performance of the machining process is strongly linked to the performance of the combination CAM system/NC unit.

Usually, CAM algorithms for tool trajectory calculation only rely on geometrical specifications and do not integrate capacities and limits of the NC unit whereas recent work pointed out that these capacities must be considered when calculating tool trajectory [1]. In particular, it is now well known that conventional machining using linear interpolators is not well adapted to high speed and highly accurate machining [1-4]. The machined surface is an approximation of the CAD surface; velocity and discontinuities appear at linear segment junctions; the amounts of data to be transmitted are very huge. These limitations among others may alter the dynamical behaviour during machining and deteriorate machining accuracy.

Therefore, recent work has concentrated on polynomial description of tool trajectories. Numerous researchers have handled the problem of real-time NURBS interpolation, with the main objective of ensuring smooth velocity all trajectory long, and/or confined chord error [2-5]. Henceforth CAM systems can calculate tool trajectories expressed as polynomial curves directly from the CAD model [6]. In the same time, Numerical Control units have evolved to interpret and control these curves.

This paper presents an analysis of the performance of the combined CAM system/Numerical Controller within the context of HSM when using native polynomial format. First, attention is focused on tool paths designed by the CAM system in terms of polynomial curves: cubic B-spline curves. It is shown that, besides the geometrical interest of native polynomial description of the tool trajectory, this format leads to a high-performance NC behaviour: efficient axis control, limitation of the slow-downs, increase of the mean feedrate …

An extension of the polynomial curve model is proposed through polynomial surfaces. Concerning this description, tool trajectories are expressed in the parametric space of the surface, which thus eliminates positioning and chord errors. Such a model allows uncoupling geometrical and dynamical effects.

The integration of surface formats can be possible via STEP-NC. We thus detail the integration of surface trajectories using STEP-NC. Our purpose is illustrated through the 3-axis machining of a free-form surface.

## 2 GENERATION OF THE TOOL TRAJECTORY

### 2.1 Tool path generation from the CAM system

Basically, the tool path is calculated from a CAD model according to a machining strategy. The machining strategy gives the machining direction and the parameters used for the CAD model approximation.

Whatever the type of machining (3 or 5 axes), the tool path can be defined by two curves, one corresponding to the trajectory of the tool extremity E, **p**(u), and the other corresponding to the trajectory of one particular point of the tool axis, D, **q**(u) (Figure 1). To remain as general as possible, we assume that **p**(u) and **q**(u) are free-form curves calculated in the part coordinate system. These curves are transmitted to the NC unit.

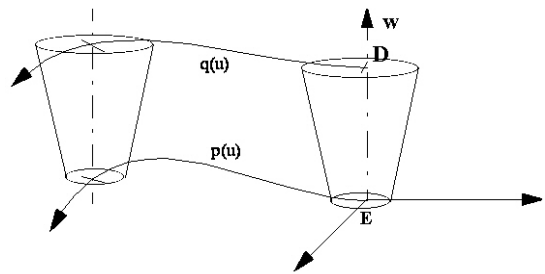

Figure 1: Description of the tool path.

### 2.2 Role and limits of the NC

The role of the NC is thus to convert the curves previously calculated into each axis' servo commands (position, velocity and acceleration). The first task of the NC motion controller can be divided into the following [2]: Interpret the machining codes coming from the CAM system; Convert the data into control commands for each sampling period; Execute servo control commands on each axis.

When using the linear interpolation format, the tool is supposed to be moving in straight line between two successive positions. Therefore, tangency discontinuities appear at the junction points. Due to the limits of the NC unit, significant decreases in feedrates compared with specified ones are observed, leading to an increase in machining time as well as an alteration of the geometrical quality of the machined surface. As an example, the interpolation cycle time that corresponds to the minimal time period to execute one block, causes the machine tool to slow-down, or stop, then reaccelerate, when the length of elementary segments becomes too small. Axis acceleration capabilities also limit the feedrate near corners or portions with high curvature [1].

To overcome all these problems, critical within the context of HSM, works have concentrated on real-time polynomial interpolation, which approximates the calculated trajectory by a polynomial trajectory. As CAM systems calculate the tool trajectory as polynomial curves, authors henceforth integrate native polynomial curves in the process of tool path NC treatment (Figure 2) [2-5]. Due to the coherence between the CAD/CAM model and the NC model of the tool trajectory, the first task of machining code interpretation is thus removed and the following tasks are simplified. This is a first step for defining an integrated model between CAD, CAM and NC treatment activities.

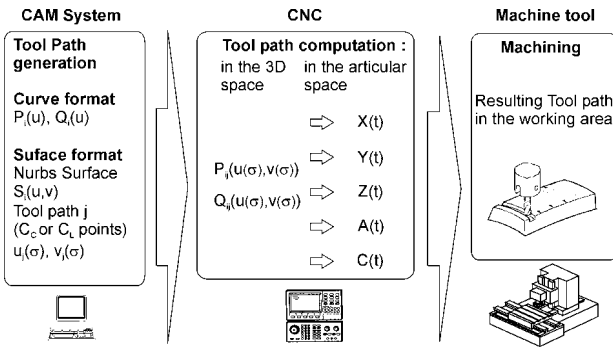

Figure 2: Method for curve machining.

Significant gains can be highlighted when using native polynomial curves as it is exposed in the next section.

## 3 NATIVE POLYNOMIAL FORMAT

This section focuses on a part of some works achieved on native polynomial formats, curves as well as surfaces, which improve performance of the couple CAM system/NC unit within the context of HSM. It should be noticed that our purpose is not to develop new real–time interpolator, nor new functionalities of NC unit, but to assess native polynomial format. Therefore, an industrial Numerical Controller (Siemens 840D) and all its functionalities has been used. In particular, this constrains the choice of the tool trajectory geometrical model: tool trajectories are defined as cubic B-spline curves.

### 3.1 Native B-spline curves

Let $\mathbf{p}(u)$ be the curve defining the trajectory of E, and $\mathbf{q}(u)$ that corresponding to the trajectory of D; $\mathbf{p}(u)$ and $\mathbf{q}(u)$ are cubic B-spline curves defined by:

$$\mathbf{p}(u) = \sum_{i=0}^{n} N_{im+1}(u)\mathbf{P}_i \; ; \; \mathbf{q}(u) = \sum_{i=0}^{n} N_{im+1}(u)\mathbf{S}_i \quad (1)$$

where $u \in [u_0, u_{n+m+1}]$ defines the knot sequence, $N_{im}$ are the basic functions for which degree m is equal to 3, and where $\mathbf{P}_i$ and $\mathbf{S}_i$ define the (n+1) control points of both B-spline curves. To ensure a consistent machining, the parameterisation of both curves must be the arc-length. In 3-axis parallel plane machining using a ball-end cutter tool $\mathbf{p}(u)$ and $\mathbf{q}(u)$ can be calculated as suggested in [7].

Each trajectory is defined in the parametric space of the offset surface as the intersection curve between the machining plane and the offset surface.

More generally, the curves are calculated to define the movement of the tool extremity and of the tool axis orientation in the part coordinate system. Both B-spline curves can be calculated according to Langeron *et al.* [6], who propose a general algorithm of calculation that can be applied for all types of machining, whatever the tool geometry, and whatever the guiding mode. An initial set of points corresponding to the initial tool path is calculated using usual positioning methods verifying the geometrical specifications. A subset of these points defines a set of interpolating points that are fitted by a cubic B-spline curve while verifying a specified interpolation error. The method can be applied to both $\mathbf{p}(u)$ and $\mathbf{q}(u)$ curve. The scheme of interpolation is a classical one. The unit vector of the tool axis is calculated, for each parameter u, from:

$$\mathbf{w}(u) = \sum_{i=0}^{n} N_{im}(u)(\mathbf{S}_i - \mathbf{P}_i) \bigg/ \left\| \sum_{i=0}^{n} N_{im}(u)(\mathbf{S}_i - \mathbf{P}_i) \right\| \quad (2)$$

When using such a method, $C^2$ continuous curves are associated to subsets of interpolating points defining tool path portions. To make the interpolation method efficient, tool path portions must be at least $C^2$ continuous that constrains to exactly identify these portions from the initial calculated tool path. This point still constitutes a difficulty [6-9].

It is important to notice that recent NC units have evolved to perform all tool path treatments in the part coordinate system, such as inverse kinematics transformation [3,6]. This justifies the sole calculation and transmission to the NC of $\mathbf{p}(u)$ and $\mathbf{q}(u)$ even in 5-axis machining (Figure 2).

First experiments show benefits when using polynomial format in terms of machining time, dynamical behaviour, and machined surface quality. A set of test parts has been designed either for 3 or 5-axis machining. To assess performance improvement of the couple CAM system/NC unit when using native polynomial trajectories, we have machined those parts on a 5-axis Mikron Machine Tool equipped with a Siemens 840D. Only a few examples are presented in Figure 3, which illustrates:

- Limitation of the slow-downs when traversing C1 discontinuity appearing near segment junction points (green curve, Figure 3a).

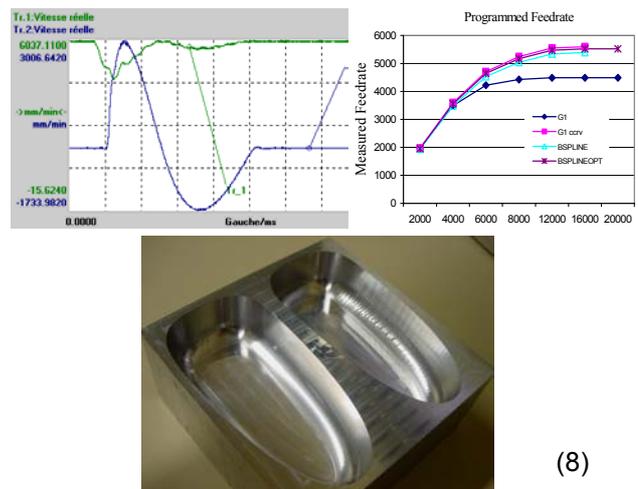

(8)

Figure 3: Illustrations of polynomial machining.

- The mean value of the feedrate is closer to the programmed one than when using G1 format (Figure 3b). In this case, the real-time interpolation has been also tested, and shows behaviour similar to that of the native polynomial format. However, if the dynamical behaviour is good high enough, the real-time interpolation leads to important geometrical errors.

- Facets are eliminated on large curvature radius (Figure 3c left) due to the smoothness of the tool trajectory.

Complete results are exposed in [6] and [9].

It should be noticed that the use of a $C^2$ continuous model allows developing optimisation techniques in order to integrate constraints linked to the dynamical behaviour, to geometrical errors, and so on. These optimisation methods are simply based on curve deformations [10].

However, this model presents some limits as for the tool positioning errors that can be removed by the use of a surface model. Therefore, the other part of our contribution concerns the generalisation of the previous model to a surface model.

### 3.2 Polynomial surface format

To generalise the polynomial format, **p**(u) and **q**(u) are curves defined in the parametric space of surfaces. Therefore, the continuity is preserved in the direction normal to the machining direction. This concept was first introduced by Duc *et al.* [10] and was developed by Tournier *et al.* [11].

Indeed, let us consider **S**(u,v) a surface to be machined in 3-axes using a ball-end cutter tool. If $S^{of}$(u,v) is the offset surface of S, by a distance equal to the tool radius R, then $S^{of}$ is the surface representation of the tool path: all tool trajectory that exactly machines the surface belongs to $S^{of}$.

This concept was extended in 5-axis machining. In flank milling, both curves **p**(u) and **q**(u) define a ruled surface corresponding to the tool trajectory. This format allows optimisation for tool positioning in order to minimize geometrical deviations [10]. Besides, deforming **q**(u) by control point displacement permits to control tool axis orientation. This point is still under work, in particular for avoiding singular positions of the work machine space or for defining optimal tool axis orientation as regards control laws of the controller.

Concerning the general case of 5-axis machining in point milling using a filleted end cutter tool, a generic definition has been suggested [11]. Two surfaces are necessary to calculate the tool trajectory. Let us consider r and R, the corner radius and the radius of the tool (Figure 4). If $C_c$ is the contact point between the tool and the surface to be machined, K is defined as the offset point of $C_c$ by a distance value equal to r. When the tool is set up in position, the point K remains fixed. Therefore, points K and $C_L$ and the normal **n** allows positioning the tool. which defines two surfaces: $S_1$, the guiding surface, locus of the K point, and, $S_2$, the orientation surface locus of the $C_L$ point, called (Figure 4).

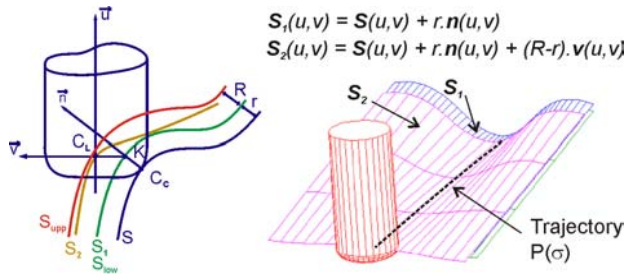

Figure 4: Definition of the surface format in 5 axes.

Both surfaces allow uncoupling dynamical and geometrical effects. As the guiding surface is independent from the machining strategy, trajectory of the tool tip is defined in the parametric space of the guiding surface as:

$\mathbf{p}(\tau) = \mathbf{S_1}(u(\tau),v(\tau))$ (3)

The orientation surface defines the tool axis orientation according to the machining strategy. $S_2$ is located between two surfaces $S_{inf}$ and $S_{sup}$, corresponding to extreme conditions. Note that $u(\tau)$ and $v(\tau)$ can also be defined as B-spline curves in the 2D parametric space.

This model turns out to be a generic model for tool path planning or tool path optimisation by deforming the guiding surface to answer geometrical criteria, or the orientation surface to answer dynamical ones [11].

The surface format imposes to manage some difficulties. The first one corresponds to the change of surface to be machined that implies the tool to pass from the parametric space of one surface to the other one. The junction of both machining curves must be managed. When the contact point is no more defined on the surface to be machined or on the tool trajectory, the tool trajectory must be prolonged by a curve defined in the 3D space. These cases generally appear when the surface presents holes, or back-drafts, but most often for outward movements of the tool as for approach, retract or link trajectories (Figure 5). The whole trajectory thus consists of a succession of parametric curves defined in the 2D parametric space of the surface and of 3D curves. As the NC unit has to treat various formats, it would become more difficult to optimise the combination Numerical Controller/Machine Tool.

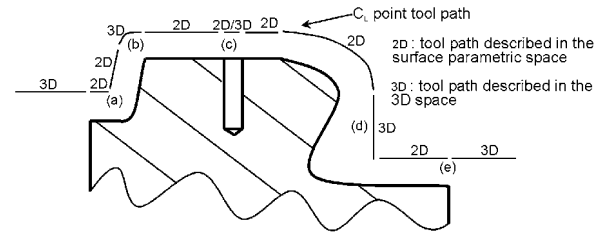

Figure 5: Tool trajectory as a succession of curves.

However, as no industrial controllers interpret such a format, real machining cannot be analysed to assess our format. Therefore, we propose to study the possible integration of surface formats through the interface STEP-NC.

## 4 EXPERIMENTATION OF SURFACE FORMATS

### 4.1 Step NC project

The European project STEP-NC showed the feasibility of an interface named STEP-NC between the CAD/CAM system and the numerical controller [12]. This is a neutral language that would replace G-codes by an object-oriented approach based on STEP. This language semantically richer concentrates on the geometry to machine, and on the machine axis movements.

In STEP-NC, tool paths are implicitly described by a feature and a strategy. The feature is a list of surfaces to machine. The strategy can be chosen among those generally used in CAM systems. In multi-axis free-form machining, it can be preferable to explicitly describe the tool path in order to control the process. Several possibilities are proposed (Figure 6). The axis coordinates of the machine tool can be sent as a curve. That means that the kinematics transformation takes place in the CAM system, and the solution is thus dedicated to a given machine, which is not a generic solution. It is also possible to define the position of a particular point of the tool ($C_C$ or $C_L$ point), either by a 3D curve, or by a pcurve defined in the 2D parametric space of the surface. The use of a pcurve seems a coherent approach as regards our previous work (section 3.2).

Therefore STEP-NC makes possible the use of a surface to describe tool trajectories: the surface model and the pcurves defining the $C_c$ or $C_L$ point trajectories are transmitted to the NC unit.

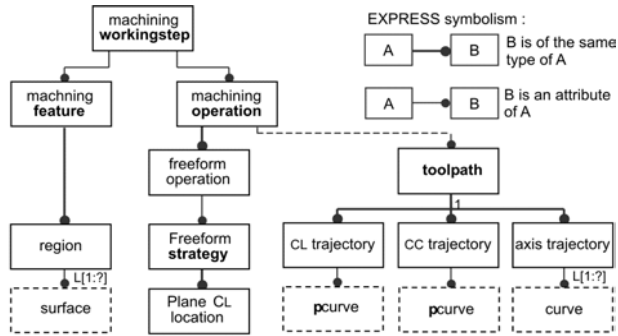

Figure 6: Free-form machining with STEP-NC.

At the present time, STEP-NC is still in the state of a project, and to assess the interest of a surface approach, we exploit interpolation format available on the NC unit.

### 4.2 Experiments

To perform our application, we define a test part that consists of portion of sphere that is linked to a plane by a tore surface. The machining strategy is supposed to be parallel planes, and the intersection between one plane and the surface to be machined, which is a pcurve consists of a part of line, a quartic, and a portion of circle (Figure 8). We retain the NURBS cubic format of the Siemens 840D (POLY curve) considering that both the line and the circle can be exactly represented by cubic NURBS. Unfortunately, the quartic must be approximated which leads to geometrical deviations. Therefore, this approximation creates a form deviation of the surface we try to minimize. Nevertheless, the *orange skin effect* observed with the linear format (Figure 7, left) is removed due to the synchronisation of the passes, synchronisation simplified by the use of quadratic rational curves to approximate the quartic. Geometrical deviations are evaluated by calculating for each point of the surface its distance to the envelope surface of the calculated tool path. With the NURBS format, errors are null on the line and the circle as expected (yellow zones, Figure 7, right), whereas they can reach 0.04mm near the quartic (red zones, Figure 7 right).

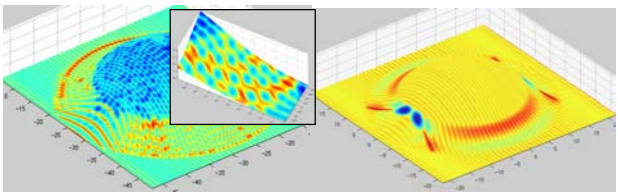

Figure 7: Geometrical deviations, (linear & NURBS).

As the dynamical behaviour is concerned, Figure 8 shows that effective feedrates are smoother than when using linear format. Slow-downs are limited, and the programmed feedrate (Vf = 4m/min) is approached.

Therefore, surface format is a good compromise between time, smoothness, and surface quality.

## 5 CONCLUSION

To conclude, polynomial formats of tool trajectories are well adapted to High Speed Machining of free-form surfaces. The use of native polynomial formats offers possibility to optimise tool paths since the calculation step by the CAM system for answering geometrical as well as dynamical constraints. Surface formats are the generalisation of polynomial formats allowing among other advantages to uncouple dynamical and geometrical effects. One polynomial trajectory is defined in the 2D parametric space of the surface. The STEP-NC interface seems a solution to integrate surface formats. The surface to be machined will in this case be transmitted to the NC unit, and curves defining tool trajectories. However, the NC treatment will require a great power of calculation to generate axis commands for those curves.

If this way seems promising since it offers a generic format for free-form machining, its integration in industrial NC units remains at the moment a perspective.

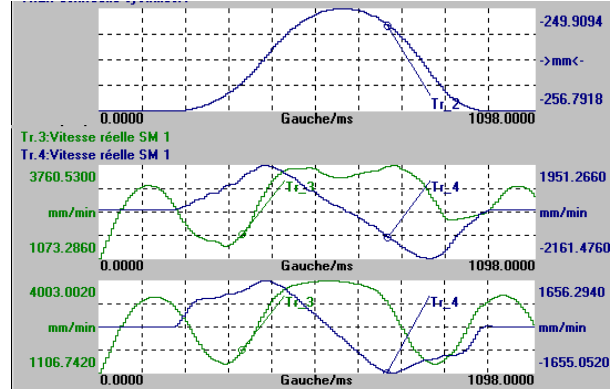

Figure 8: Feedrate analysis.